\documentclass{IAU-TrA}
\usepackage{graphicx}

\title[WORKING GROUP ON COLLISION PROCESSES]
{}

\author[DIVISION~B / COMMISSION~14 / WORKING GROUP 
ON COLLISION PROCESSES]
{}

\pubyear{2015}
\volume{Volume XXIXA}
\pagerange{001--008}
\jname{Transactions IAU, Volume XXIXA}
\editors{Thierry Montmerle, ed.}
\begin{document}

\maketitle

{\bf

\large
\noindent
DIVISION B / COMMISSION~14 / WORKING GROUP                       \\
COLLISION PROCESSES \\

\normalsize

\begin{tabbing}
\hspace*{45mm}  \=                                                   \kill
CO-CHAIRS       \> Gillian Peach                                     \\
                \> Milan S. Dimitrijevic                             \\
                \> Paul S. Barklem                                       \\
\end{tabbing}

\noindent
TRIENNIAL REPORT 2012-2015
}

\firstsection 

\section{Introduction}

Since our last report (\cite{Ped12}), a large number of new publications
on the results of research in atomic and molecular collision processes 
and spectral line broadening have been published. Due to the limited
space available, we have only included work
of importance for astrophysics. Additional relevant papers, not included 
in this report, can be found in the databases
at the web addresses provided in Section 6. Elastic and 
inelastic collisions between electrons, atoms, ions, and molecules are
included, as well as charge transfer in collisions between heavy particles
which can be very important.

Numerous conferences on collision processes and line broadening have been
held since our last report.  Some important 
international meetings that publish proceedings where additional data 
and their sources can be found are:

The 21$^{th}$ {\it International Conference on Spectral Line Shapes} (ICSLS)
(\cite{Devd12}), the 22$^{th}$ ICSLS (\cite{Pari14}), the 8$^{th}$
{\it Serbian Conference on Spectral Line Shapes in Astrophysics} (SCSLSA)
(\cite{Popo11}), the 9$^{th}$ SCSLSA (\cite{Popo14}), the XXVII 
{\it International Conference on Photonic, Electronic, and Atomic Collisions}
(ICPEAC) (\cite{Will12}), the XXVIII ICPEAC (\cite{Guoq14}), the 23$^{rd}$
{\it International Conference on Atomic Physics} (ICAP) (\cite{Duli13}), 
the 8$^{th}$ {\it International Conference on Atomic and Molecular
Data and their Applications} (ICAMDATA) (\cite{Gill13}), 1$^{st}$ {\it Spectral 
Line Shapes in Plasmas Code Comparison Workshop} (SLSP) (see review 
in \cite{Stam13}) and the 2$^{nd}$ (SLSP) (\cite{Stam14}). Selected papers from
the 9$^{th}$ ICAMDATA are in press in Physica Scripta.

\section{Electron collisions with atoms, ions, molecules and molecular ions}

Collisions of electrons with atoms, molecules and atomic and molecular ions 
are the major excitation mechanism for a wide range of astrophysical environments.
In addition, electron collisions play an important role in ionization and
 recombination, contribute to cooling and heating of the gas, and may contribute 
to molecular fragmentation and formation. In the following sections we summarize
recent work on collisions for astrophysically relevant species, including elastic
scattering, excitation, ionization, dissociation, recombination and electron 
attachment and detachment.

\subsection{Electron-atom scattering}

Extensive large-scale studies of elastic, excitation and ionisation cross sections
for N (\cite{2014PhRvA..89f2714W}), F (\cite{2014PhRvA..89e2713G}) and Ne 
(\cite{2012PhRvA..85f2710Z}) have been presented.  
Elastic scattering has been studied for fine-structure resolved states of Cl 
(\cite{2013PhRvA..87b2703W}) as well as for Ca (\cite{2014CaJPh..92..206H}), Mn 
(\cite{2013PhRvA..88d2706D}), In, Tl, Ga and At (\cite{2012JPhB...45d5201F}), 
and Eu and Yb (\cite{2013JTePh..58.1749K}).  New work on excitation includes: H
(\cite{2014ApJ...780....2V}), He in an electric field (\cite{2013JETP..116...48S}),
C including ionisation (\cite{2013PhRvA..87a2704W}), Ar (\cite{2014PhRvA..89b2706Z}),
Zn (\cite{2012PhRvA..86b2710D}), Cs (\cite{2014PhRvA..89c2712B}), Yb
(\cite{2012PhRvA..86b2710D}), and Pb (\cite{2013JPhB...46c5202Z}).  

\subsection{Electron-ion scattering}

Several large-scale efforts to calculate excitation data for iso-electronic sequences
have been undertaken.  For He-like ions data is now available covering most ions 
from Mg$^{10+}$ to Kr$^{34+}$ (\cite{2012PhyS...85f5301A, 2012PhyS...85b5305A,
2012PhyS...85b5306A, 2012PhyS...86c5302A, 2013PhyS...87e5302A, 2013PhyS...87d5304A}).

For Li-like ions data has been 
calculated for ions from Mg$^9+$ to Ni$^{25+}$ (\cite{2012ADNDT..98.1003A,
2013ADNDT..99..156A}).  For Be-like ions data has been presented for ions from B$^+$
to Zn$^{26+}$ (\cite{2014A&A...566A.104F}), and also Cl$^{13+}$, K$^{15+}$ and 
Ge$^{28+}$ (\cite{2014PhyS...89l5401A}) and Ti$^{18+}$ (\cite{2012PhyS...86e5301A}).
For B-like ions data covering from C$^+$ to Kr$^{31+}$ (\cite{2012A&A...547A..87L}),
and also Al$^{9+}$ (\cite{2014MNRAS.438.1223A}), have been calculated.  For Mg-like
ions data covering from Al$^+$ to Zn$^{18+}$ have been presented
(\cite{2014A&A...572A.115F}).  

Other new work on specific ions includes: $^3$He$^+$ and other one-electron ions
(\cite{2014ApJ...788...69B}), O$^{2+}$ (\cite{2012MNRAS.423L..35P,
2014MNRAS.441.3028S}), O$^{4+}$ and O$^{5+}$ (\cite{2013MNRAS.436.1452E}), Ne$^{4+}$ 
(\cite{2013MNRAS.435.1576D}), Mg$^{2+}$ and Al$^{3+}$ (\cite{2014AdSpR..54.1203E}), 
Mg$^{5+}$ (\cite{2012A&A...548A..27T}), Mg$^{7+}$ (\cite{2013A&A...556A..24G}),
Si$^+$ (\cite{2014MNRAS.442..388A}), Si$^{6+}$ (\cite{2014ApJ...787....2S}), 
Si$^{7+}$ (\cite{2012A&A...541A..61T, 2013A&A...556A..32L}), S$^{2+}$
(\cite{2012ApJ...750...65H, 2014ApJ...780..110G}), S$^{14+}$ and S$^{15+}$ 
including recombination (\cite{2012ApJ...754...86M}), Cl$^{2+}$ 
(\cite{2012ApJS..202...12S}), Ca$^{13+}$ (\cite{2012PASJ...64..131D}), Sc$^{+}$
(\cite{2012MNRAS.424.2461G}), Fe$^{2+}$ (\cite{2014ApJ...785...99B}), Fe$^{6+}$
(\cite{2014ApJ...788...24T}), Fe$^{7+}$ (\cite{2014A&A...570A..56D}), Fe$^{13+}$
(\cite{2014MNRAS.445.2015A}), Ni$^{13+}$ (\cite{2012ADNDT..98..779W}), Ni$^{17+}$ 
(\cite{2012A&A...537A..12H}), Ge$^{+}$ (\cite{2014JPhB...47v5204S}), Nb$^{+11}$
and Mo$^{+12}$ (\cite{2014ADNDT.100.1059L}), Sn$^{13+}$ including ionization 
(\cite{2014PhRvA..89d2704L}), W$^{3+}$ including ionisation 
(\cite{2013JPhB...46e5202B}), and Au$^{51+}$ (\cite{2014ChPhB..23k3401F}).

\subsection{Electron-molecule and electron-molecular ion scattering}

For molecules, new work includes: H$_2^+$ including proton-production and 
dissociative ionisation (\cite{2013PhRvA..88f2709Z}), H$_2$ including excitation
and dissociation (\cite{2012JPSJ...81j4501L, 2013ChPhL..30g3401W, 
2013ApJS..204...18K}), excitation of HD$^+$ (\cite{2014PhRvA..90a2706M}), 
dissociative ionisation of D$_2$ (\cite{2013PhRvA..88f2705L}),  dissociative 
recombination of LiH$_2^+$ (\cite{2014PhRvA..89e0701T}), excitation of BeH$^+$
(\cite{2012EPJD...66...31C, 2013PhRvA..87b2713N}),
elastic, inelastic and total cross sections for HCN (\cite{2012JChPh.137l4103S}),
elastic and inelastic scattering on H$_2$CN (\cite{2014PhRvA..89b2711W}),  
dissociative recombination of N$_2^+$ (\cite{2014PhRvA..90e2705L}), excitation
and dissociation of N$_2$ (\cite{2012PhRvA..85f2704M, 2014PSST...23f5002L,
2014PlST...16..104X}), excitation of H$_2$O (\cite{2013PhRvA..88a2703R}),
dissociative ionisation of HDO$^+$ (\cite{2014PhRvA..90d2704D}), elastic and 
inelastic scattering on CO (\cite{2012PSST...21d5005L, 2012CP....405...16R,
2013ChPhB..22b3402W}), dissociative recombination of HCO$^+$ and HOC$^+$
(\cite{2012PhRvA..85d2702L}),
elastic and inelastic scattering on NO (\cite{2013PhRvA..88b2708C}), total 
scattering and excitation cross sections for N$_2$O (\cite{2012JChPh.137g4311V,
2014JPhB...47o5203W}), elastic, excitation and ionisation cross sections for
NCO (\cite{2012PhRvA..85e2701K}), elastic, excitation and ionisation processes 
on O$_2$ (\cite{2014PhRvA..90b2714S}), fragmentation of CO$_2^{2+}$ 
(\cite{2014PhRvA..90f2705W}), excitation of NF$_3$ (\cite{2013PhRvA..88c2707G}), 
and elastic scattering on OCS and CS$_2$ (\cite{2013JChPh.138e4302M}).

Regarding more complex molecules, new references include: dissociative electron
attachment on acetone (\cite{2014JChPh.141p4320P}), excitation of allene 
(\cite{2013PhRvA..87f2701B}), elastic and inelastic scattering on formamide
(\cite{2012PhRvA..85a2706W}), dissociation of methane (\cite{2012JChPh.137vA510Z}),
elastic and inelastic scattering on pyrazine (\cite{2012JChPh.137t4307P, 
2013JChPh.139r4310S}), elastic and inelastic scattering on pyrimidine
(\cite{2012JChPh.136n4310M}), and inelastic scattering on uracil
(\cite{2012JChPh.137r4303S}).

\section{Heavy particle collisions}

Collisions between heavy particles are important in many astrophysical 
environments, particularly those involving H and He as they are astrophysically
abundant.  In the following sections we attempt to summarise recent work on
collisions for astrophysically relevant species, including excitation, ionisation
and charge transfer.  We note that a review of reactive scattering and chemistry
relevant to astrophysics is beyond the scope of the present report.

\subsection{Atom-atom and ion-atom collisions}

Excitation and charge transfer in collisions of neutral atoms with hydrogen 
atoms are important processes in stellar atmospheres, and studies include:
H + H and Be + H (\cite{2014JPhB...47v5206M}), Mg + H
(\cite{2012PhRvA..85c2704B,2012A&A...541A..80B}), Al + H 
(\cite{2013A&A...560A..60B,2013PhRvA..88e2704B}), Si + H
(\cite{2014A&A...572A.103B}), and Cs + H (\cite{2014PhRvA..90f2701B}). 
Excitation and charge transfer in H + H$^+$ collisions have been studied
(\cite{2012MNRAS.422.2357T}) and proton-Rydberg atom (\cite{2012ApJ...747...56V}),
atom-Rydberg atom (\cite{2013A&A...552A..33S}) collisions studied generally.
Charge transfer, excitation and ionisation in collisions between ions and neutral 
atoms play important roles in a number of astrophysical environments.  New work 
on processes involving hydrogen atoms and protons include: excitation, ionisation
and charge transfer in collisions between protons and He$^+$, Li$^{2+}$, 
Be$^{3+}$, B$^{4+}$, and C$^{5+}$ (\cite{2013PhRvA..87c2704W}), proton collisions
with He (\cite{2012JPhB...45v5203L}), excitation and charge transfer in
Li$^{q+}$ + H collisions (\cite{2014PhPl...21f2513L}), charge transfer in
Be$^{3+}$ + H collisions (\cite{2013PhRvA..87d2709L}), charge transfer and 
ionisation of Be, Fe, Mo and W by H collisions (\cite{2014JPhB...47c5206T}), 
charge transfer in C$^{5+}$ + H collisions (\cite{2012JPhB...45x5202N}), 
charge transfer and ionisation in C$^{6+}$, N$^{7+}$ + H collisions 
(\cite{2014EPJD...68..227J}), charge transfer in O$^{6+}$ + H collisions 
(\cite{2012JPhB...45w5201W}), charge transfer and ionisation in O$^{8+}$ + H 
collisions (\cite{2012PhPl...19f2104P}), charge transfer and ionisation in 
He-like ions (Li$^{+}$, Be$^{2+}$, B$^{3+}$, C$^{4+}$, N$^{5+}$, O$^{6+}$) + H
collisions (\cite{2013PhPl...20b2104P}), charge transfer and ionisation in 
N$^{7+}$, N$^{6+}$, C$^{6+}$ + H collisions (\cite{2012JPhB...45f5203I}), 
charge transfer in Si$^{3+}$ + H collisions (\cite{2014PhRvA..90a2708L}), 
charge transfer in Kr$^{36+}$, W$^{60+}$ + H collisions 
(\cite{2013PhST..156a4033I}) and charge transfer of W$^+$ and W$^{2+}$ with H 
and H$_2$ (\cite{2012JPhB...45n5201T}).  

New references regarding charge transfer involving He and He$^+$ include:
He$^+$ + He (\cite{2012PFR.....701062K}), Li + He (\cite{2014A&A...565A.106B}),
Si + He$^+$ (\cite{2013MNRAS.436.2722S}), C$^{4+}$ + He collisions
(\cite{2013PhRvA..88b2706Y}), O$^{8+}$ + He collisions 
(\cite{2012CaJPh..90..283W}), Ne$^{10+}$ + He, Ne (\cite{2014PhRvA..89a2710L}).
An overview of charge transfer and ionisation for heavy many-electron ions 
colliding with neutral atoms has been given in (\cite{2013PhyU...56..213T}) and
W$^+$, W$^{2+}$ + He (\cite{2012JPhB...45n5201T}).  A classical model for ion
collisions with He atoms, including charge transfer and ionisation processes,
has been presented (\cite{2012PhyS...85a5302D}).  Other references include
studies of charge transfer of Ar with highly charged ions
(\cite{2014PhRvA..90f2708O}), Ar$^{16+}$ + Ne (\cite{2014PhRvA..90e2720X}).   

The study of polarisation in spectral lines requires information on the 
destruction of polarisation by collisions, including with H.  Some recent 
studies include those of \cite{2012A&A...545A..11D}, \cite{2014A&A...572A..53D}
and \cite{2014ApJ...788..118M}.\\

\subsection{Atom-, ion-, and molecule-molecule Collisions}

A second edition of the book {\it Molecular Collisions in the Interstellar
Medium} (\cite{2012mcim.book.....F}) has been published and also 
\cite{2013ChRv..113.8906R} have published a review entitled 
{\it Molecular Excitation in the Interstellar Medium: Recent Advances in
Collisional, Radiative and Chemical Processes}.
 
For excitation, new references include:  CO + H$^+$, H$_2^+$, 
H$_3^+$ (\cite{2014ApJ...780..157W}), CO + H (\cite{2013ApJ...771...49Y}),
HD + He (\cite{2012ApJ...744...62N}), OH$^+$ + He (\cite{2014ApJ...794...33G}),
HCl + He (\cite{2012MNRAS.424.1261L, 2014ApJ...783...92Y}), H$_2$O + He 
(\cite{2013ApJ...765...77Y}),   DCO$^+$ + He (\cite{2012MNRAS.421..719B}),
C$_2$H + He (\cite{2012MNRAS.421.1891S}), H$_2$CO + He 
(\cite{2014AdSpR..54..252S}), H$_2$ + H$_2$ (\cite{2014JChPh.140f4308B}),
C$^+$ + H$_2$ (\cite{2014ApJ...780..183W}), CN + H$_2$ 
(\cite{2012MNRAS.422..812K}), O$_2$ + H$_2$ (\cite{2014A&A...567A..22L}), 
HCO$^+$ + H$_2$ (\cite{2014MNRAS.441..664Y}), NH$_2$D + H$_2$ 
(\cite{2014MNRAS.444.2544D}).

For charge transfer and related processes, new work includes: H$^+$, H$_2^+$, 
H$_3^+$ + CO (\cite{2014ApJ...780..157W}), charge transfer involving multiply
charged ions and CO (\cite{2013PhRvA..88a2714C}) charge transfer of 
He$^+$ with simple molecules (\cite{2012PFR.....701062K}), single and multiple
electron charge transfer and ionisation in H$^+$+H$_2$O 
(\cite{2012PhRvA..85e2704M}), as well as multiple electron removal and 
fragmentation processes in He$^+$+H$_2$O collisions (\cite{2012PhRvA..86b2719M}).

\section{Stark broadening}

Stark broadening parameters are required
for the analysis, synthesis and interpretation of stellar spectra, and for other
astrophysical plasmas, such as H I and H II regions or neutron stars.
Data on line widths and shifts for a large number of atomic transitions are 
particularly important for hot dense stars such as white dwarfs,
where Stark broadening is usually the most important broadening mechanism.

\subsection{Developments in line broadening theory}

\cite{Stam13a} have developed an analytical method based on the quasicontiguous
approximation to the static line shapes for Stark broadening
of H and H-like transitions including Rydberg ones. 
The unified formalism for the modeling of hydrogen Stark line shapes has been 
reexamined by \cite{Rosa13} and extended to non-binary interactions between an 
emitter and the surrounding perturbers.  
\cite{Oks15} has reexamined conventional theory for Stark 
broadening, and his new calculations for the H$_{\alpha}$ line show that 
the effect of the ion dynamics may be slightly smaller than was 
accepted, while the effect of the acceleration of perturbing electrons by
the ion field in the vicinity of the radiating atom may be larger.

\cite{Alex13} has proposed the implementation in lineshape codes of the
Frequency Separation Technique based on the separation of the slow and fast
frequency components.

Kinetic quantum equations for a density matrix with collision integrals 
describing nonlinear effects in line wing spectra, have been derived by 
\cite{Park11}. \cite{Rosa12} have investigated the influence of correlated
emitter-perturber collisions on hydrogen lines using kinetic theory. 
They concluded that such collisions strongly affect the width and shape of
the line profile in the core region where the dynamical effects are included using
the frequency-fluctuation-model approach. \cite{Demu14} have studied the assumption
that the density matrix is diagonal for the calculation of spectral lineshapes
and the effect of microfield rotation on Stark profiles. 
The electron-impact broadening of ion lines has been calculated by \cite{Naam14}
using trajectories modified by non-Newtonian mechanics and the Maxwell-J\"uttner
velocity distribution.

The effect of the microfield direction on line shapes has been investigated by
\cite{Cali14} and the ion-dynamic effect has been also analyzed by \cite{Ferr14}.
A stochastic microfield formulation to treat the effect of particle motion
in the Stark broadening of ion lines is extended by \cite{Igle13} to include an
external uniform static magnetic field.  Efficient algorithms for calculations
of Stark-Zeeman line profiles for both static and dynamic ions are presented. 
\cite{Chen11} have investigated the influence of a non-uniform microfield
on the asymmetry of Ly$_{\alpha}$ in a dense plasma and \cite{Difa12} have 
studied the effect on its profile of an oscillating electric field.

\cite{Bacl13} have presented a new approach for the evaluation of asymmetry 
parameters, and \cite{Galt13} have investigated quantum mechanical 
interference effects in Stark broadening of intrashell transitions for dense
plasma conditions.
Problems arising from the ideal Coulomb plasma approximation have been analysed
by \cite{Rosa14} and \cite{Sapa12} have derived new analytical series 
expansions for the hydrogenic line shape functions.

\cite{Saha14a} have reviewed and analyzed the semiclassical perturbation theory
(SCP) in order to introduce the main approximations, give the formulae used
in the SCP code and to better understand the validity conditions.
A review of different models used for the calculation of Stark broadening 
is presented by \cite{Gigo14a}.

\subsection{Isolated lines}

Stark broadening of isolated lines is dominated by plasma-electron impacts. 
Broadening parameters, Stark widths and shifts, have been determined
theoretically for:

2 B III (\cite{Duan14}), 8 Si IV (\cite{Elab12}) 6 Ar XV (\cite{Elab14}) lines
and for 7 Sr II lines in an ultracold neutral plasma (\cite{Duan13}), using a
quantum-mechanical approach in all cases.

For 9 resonant Cr II multiplets (\cite{Simi13}), 114 Pb IV lines (\cite{Hamd13}),
32 Ar III lines (\cite{Hamd14}), 148 C II 
multiplets (\cite{Larb12}), and 36 B IV (\cite{Dimi14}), new Stark broadening
parameters are calculated using a semiclassical perturbation approach. A 
semiclassical method is also used for the N I ($^1$D)3s$^2$D -
($^1$D)3p$^2$D$^o$ doubly excited transition.

The Li I 460.28 nm spectral line and its forbidden component has been 
investigated using computer simulation (\cite{Cvej14}).

A semi-empirical approach, that uses Hartree-Fock relativistic wave 
functions and includes core
polarization effects, has been applied to 237 Mg III (\cite{Colo13}),
111 Mg III (from configurations 2p$^5$4f, 2p$^5$5f and 2p$^5$5g - 
\cite{More14}), 148 Ca III (\cite{Alon13}), 467 Ca IV (\cite{Alon14}), and 72 
lines of Pb V (\cite{Alon12}). A modified semi-empirical approach is used for 
15 Nb III (\cite{Simi14}) lines. 

Broadening parameters have been obtained experimentally for the
following numbers of lines:

Li I 460.28 nm
line with forbidden component (\cite{Cvej14}), 6 He I 
(\cite{Gigo14}), 3 Mg I and 1 Mg II (\cite{Cvej13}), 1 N I (\cite{Bart14}), 
4 Al II (\cite{Ciri14}), 126 Ar II (\cite{Djur13}), 
13 Ar II (\cite{Gajo13}), 11 Ca II (\cite{Agui14a}), 26 Fe II (\cite{Agui11}), 
36 Fe II and 27 Ni II 
(\cite{Arag14a, Arag14b}), 2 Cu I (\cite{Burg14b}), 22 Cu I and 
100 Cu II (\cite {Skoc13}),
1 Hg I, 19 Hg II, 6 Hg III, and 4 Hg IV (\cite{Gavr12}), 2 In I (\cite{Burg14a}), 
4 In III (\cite{Skoc12}), 53 Ni II (\cite{Arag13}), and 83 Cr II 
(\cite{Agui14b}) spectral lines.

Regularities and systematic trends of Stark broadening parameters
within various spectral series have been investigated  for spectral
lines of helium (\cite{Dojc12}), lithium (\cite{Dojc13b}), beryllium
(\cite{Dojc11}), calcium \cite{Tapa12}), potassium (\cite{Jevt12}), 
and within homologous spectral series of alkaline earth metals 
(\cite{Dojc13a}). The systematic trends obtained and data on regular 
behaviour within spectral series, can be of interest for the
spectroscopic diagnostic of astrophysical plasmas, since this enables 
the estimation and prediction of missing Stark broadening parameters.

\subsection{Transitions in hydrogenic and helium-like systems}

The Stark-broadened profile of Ly$_{\alpha}$ has been obtained using {\it ab initio}
simulation calculations by generating the electric microfield with a renewal
process (\cite{Hamm12}). Also Stark broadening of H$_{\beta}$ has been
investigated for cold plasmas with low electron densities (\cite{Palo12}).  

New relativistic quantum mechanical calculations of Stark widths and shifts
of He II spectral lines have been published (\cite{Duan12}).

The Stark broadening of He I lines 501.6, 667.8, 728.1, 388.9, 587.6, and 
706.5 nm has been studied theoretically and experimentally (\cite{Gigo14}). 
The theory uses computer simulations of the dynamics for noninteracting
particles.
Also broadening of He I 492.2 nm line has been investigated theoretically
by computer simulation (\cite{Lara12}), and with its forbidden components 
experimentally (\cite{Ivk13}).  New Stark broadening parameters have been
calculated for the allowed 447.1 nm line and its forbidden component at 447.0 nm
(\cite{Faye11}), as well as for the 388.9 nm line (\cite{Omar14}).

New calculations of complete Stark broadened profiles for 15 radio recombination
lines of hydrogen with the principal quantum number of the upper level n between
102 and 274, have been performed recently (\cite{Peac14}).  Electron and proton 
impact has been included and it has again been confirmed that for the radio
recombination lines electron-impact broadening is the dominant broadening mechanism.

\section{Broadening by neutral atoms and molecules}

The analysis of experimental molecular spectra in order to extract line
shape parameters is often very difficult.  Line shapes can be affected
by collisional narrowing and the dependence of collisional broadening
and shifting on molecular speed.  When these effects are sufficiently
important, fitting Voigt profiles to experimental spectra produces
systematic errors in the parameters retrieved.  Several theoretical
papers have been published that address these problems of modelling
and computation, see \cite{Koch11}, \cite{Stac12}, \cite{West12},
\cite{Koch13}, \cite{Berk13}, \cite{Prot13}, \cite{Wcis13},
\cite{Tran13} and \cite{Wang14}.\\

The applicability or otherwise of power-law relations to describe the
temperature variations of the pressure broadening and shifts of atomic
and molecular lines have been studied by \cite{Cybu13}.\\

In the following sections the experimental (E) and theoretical (T) 
results selected have been confined to the basic atomic and molecular 
data required for a description of the pressure broadening and shift
of lines and molecular bands.

\subsection{Broadening and shift of atomic lines}

Some new research has been published in the period 2012-2015 and the
transitions studied together with the perturbing atoms or molecules
are listed below.
\\[1ex]
He: self-broadening of 1s--2p and 2p--3s transitions (T) (\cite{Alla13}).\\
Li: broadening of lines by He (T) (\cite{Alla14b}).\\
Li: broadening of lines by H$_2$ (T) (\cite{Alla14a}).\\
Na: broadening of lines by H$_2$ (T) (\cite{Alla12}).\\
K, Rb, Cs: broadening, shift and asymmetry of D$_1$ and D$_2$ lines by
   He, Ne, Ar (T) (\cite{Blan14}).\\
K, Rb: D$_1$ and D$_2$ lines broadened and shifted by ${}^3$He, N$_2$ (E)
   (\cite{Klut13}).\\
K: D$2$ line broadened and shifted by He, Ne, Ar, Kr, Xe, N$_2$, CH$_4$,
   C$_2$H$_6$, C$_3$H$_8$, n-C$_4$H$_{10}$ (E) (\cite{Pitz14}).\\
Rb: D$1$ line broadened and shifted by He, CH$_4$, C$_2$H$_6$, C$_3$H$_8$,
   n-C$_4$H$_{10}$ (E) (\cite{Pitz14}).\\
Ca$^+$: broadening of lines by He (T) (\cite{Alla14c}).\\
Cd: blue and red wings of the 326.1nm line perturbed by He, Ne, Ar, Kr, Xe 
   (E) (\cite{Rost14}).\\ 
Ag: D1 line broadened by He, Ar and N$_2$ (E) (\cite{Kara12}).\\[1ex]
Dispersion coefficients C$_6$, C$_8$ and C$_{10}$ for interactions between
H, Li, Na and K atoms (T) (\cite{Kar13}).\\
Broadening of radiative transitions in charge exchange collisions of 
one-electron atomic ions and bare nuclei, also H$^-$ + H (T) (\cite{Devd14}).

\subsection{Broadening and shift of molecular lines}

Much new data has been published since the last report was prepared.
The molecules are listed below with their perturbing atomic or
molecular species.
\\[1ex]
H$_2$-H$_2$: electric quadrupole transitions and collision-induced
   absorption (E) (\cite{Kass14}).\\
H$_2$-H$_2$: collision-induced absorption (E) (\cite{Abuk15}).\\[1ex]
H$_2$O: lines broadened and shifted by H$_2$ (E+T) (\cite{Drou12}).\\
H$_2$O: broadened by He isotopes (E) (\cite{Camp13}).\\
H$_2$O: lines broadened by air (E) (\cite{Birk12}).\\
H$_2$O: intermolecular potentials; vibration bands broadened by He (T)
   (\cite{Petr13}).\\
H$_2$O: self-broadened vibration-rotation line (E+T) (\cite{DeVi11}).\\ 
H$_2$O: self-broadened lines and broadened by air (E) (\cite{Ngo12}).\\
H$_2$O: self-broadened lines, pressure-induced shifts (T) (\cite{MaQ12}).\\
H$_2$O: self-broadened water vapour continuum (T) (\cite{Klim13}).\\
H$_2$O: self-broadened water vapour continuum (E) (\cite{Ptas13}).\\
H$_2$O: water vapour continuum broadened by air (E) (\cite{Sloc13}).\\
H$_2$O: self-broadening coefficients for lines of H$_2^{16}$O, H$_2^{17}$O,
   H$_2^{18}$O, HD$^{16}$O (E) (\cite{Rega14}).\\
H$_2$O: rotational band broadened by N$_2$ (T) (\cite{Lamo12a}).\\[1ex]
H$_2$O, CO: lines broadened by H$_2$ (T) (\cite{Faur13}).\\
H$_2$CO: anomalous absorption in molecular lines (E) (\cite{Shar12}).\\
HO$_2$: lines broadened by N$_2$, O$_2$ (E) (\cite{Mizo12}).\\
HO$_2$: $\nu_3$ band broadened by air (E) (\cite{Mina14}).\\
HDO: self-broadened lines (E) (\cite{Daum12}).\\
(H$_2$O)$_2$: lines perturbed by H$_2$O (T) (\cite{Odin14}).\\
HCl, HBr: lines broadened and shifted by N$_2$ (E) (\cite{Asfi12}).\\[1ex]
CH$_4$: self-broadened lines (E) (\cite{Bowl12}), (\cite{Cerm12}).\\
CH$_4$: self-broadened lines and broadening by N$_2$ (E) (\cite{Sanz12}).\\
CH$_4$: lines broadened and shifted by CH$_4$ and air (E) (\cite{Smit14}).\\
CH$_4$: collision-induced absorption (T) (\cite{Elka12}).\\
CH$_4$: lines broadened by CO$_2$ (E) (\cite{Fiss14b}).\\
CH$_4$: lines broadened by H$_2$, N$_2$, O$_2$ (T) (\cite{Gaba13}).\\
CH$_4$: lines broadened and shifted by N$_2$ (E) (\cite{Kapi12}), (\cite{Visp14}).\\
CH$_4$: line mixing and lines broadened by air (E) (\cite{Ghys14}).\\
CH$_4$: lines broadened by N$_2$, Ne (E) (\cite{Kapi13}).\\[1ex]
CH$_3$D: lines broadened and shifted by N$_2$, O$_2$, CO$_2$ (E) (\cite{Tang13}).\\
CH$_3$F: self-broadened lines (E) (\cite{Okab15}).\\
CH$_3$Br: self-broadened lines and broadened by N$_2$ (E) (\cite{Bous15}).\\
CH$_3$Cl: self-broadened lines and broadened by N$_2$ (E+T)
   (\cite{Bray12}), (\cite{Bray13a}), (\cite{Bray13b}).\\
CH$_3$Cl: lines broadened by O$_2$ (E+T) (\cite{Buld13}).\\
CH$_3$Cl: self-broadened lines (E+T) (\cite{Ramc13}), (\cite{Ramc14}).\\
CH$_3$Cl: lines broadened by CO$_2$ (E) (\cite{Duda14}).\\
C$_2$H$_2$: lines self-broadened and shifted (T) (\cite{Gala13a}).\\
C$_2$H$_2$: lines broadened by N$_2$ (E+T) (\cite{Hash14}).\\
C$_2$H$_2$: self-broadened lines and broadening by N$_2$, He, Ar (E)
   (\cite{Saji14}).\\
C$_2$H$_4$: self-broadened lines (E) (\cite{Auwe14}).\\
CH$_3$OH: lines broadened by air (E) (\cite{Harr12}).\\
CH$_2$CCH$_2$: self-broadened lines and broadened by N$_2$ (E) (\cite{Fiss13}),
   (\cite{Fiss14a}).\\[1ex]
CO, CO$_2$: lines broadened by H$_2$ (E) (\cite{Padm14}).\\
CO: self-broadened lines and broadened by air (E) (\cite{Devi12a}),
   (\cite{Devi12b}).\\
CO$_2$: self-broadened lines (E) (\cite{Deli12}).\\
CO$_2$: self-broadened lines (T) (\cite{Hart13b}).\\
CO$_2$: absorption coefficients at high temperature and Pressure (E)
   (\cite{Stef13}).\\
CO$_2$: lines broadened and shifted by N$_2$ (T) (\cite{Gama12}).\\
CO$_2$: lines broadened and shifted by O$_2$, air (T) (\cite{Lamo12b}).\\
CO$_2$: lines broadened and shifted by N$_2$, O$_2$, CO$_2$, air (T)
   (\cite{Gama13}), (\cite{Gama14}).\\
CO$_2$, C$_2$H$_2$, CO, HCl, HF: lines broadened by Ar (T) (\cite{Ivan13}).\\[1ex]
Cs$_2$: lines broadened by He, Ne, Kr (\cite{Tshi12}), (\cite{Tshi14}).\\
N$_2$: self-broadened lines (T) (\cite{Thib12}).\\
O$_2$: self-broadened lines (T) (\cite{Hart13a}).\\
O$_2$: lines broadened by air (E+T) (\cite{Lamo14}).\\
O$_2$: lines broadened by H$_2$O (E) (\cite{Drou14}).\\
OCO: self-broadened band lines (E) (\cite{Ngo14}).\\
OCS: bands broadened and shifted by O$_2$ (E) (\cite{Gala13b}).\\[1ex]
SO$_2$: lines broadened by H$_2$, N$_2$, O$_2$, He (E) (\cite{Cazz12}).\\
NH$_3$: self-broadened lines (T) (\cite{Cher14}).\\
NH$_3$: self-broadened isolated rovibrational line (E+T) (\cite{Trik12}).\\
NH$_3$: lines broadened by N$_2$, O$_2$, CO$_2$, H$_2$O (E) (\cite{Owen13}).\\
PH$_3$: lines self-broadened and shifted (E) (\cite{Devi14}).\\
PH$_3$: perturbed by $H_2$, line-mixing coefficients (E) (\cite{Sale14}).\\

\section[Databases]{Databases}

The previous list of useful databases (\cite{Ped12}) is updated, changes
of web adresses are checked and the corresponding new references added. The 
selected databases are:

Vienna Atomic Line Database (VALD) is a collection of atomic and molecular 
transition parameters of astronomical interest for the analysis of
radiation from astrophysical objects, containing central wavelengths,
energy levels, statistical weights, transition probabilities and line
broadening parameters for all chemical elements of astronomical importance.
The newest version VALD3 can be found at http://vald.astro.uu.se/
(\cite{Kupk99, Kupk11}).

CHIANTI database (\cite{Dere09, Land12, Land13}) contains a critically 
evaluated set of
up-to-date atomic data for the analysis of optically thin collisionally
ionized astrophysical plasmas.  It lists experimental and calculated 
wavelengths, radiative data and rates for electron and proton collisions,
see \\ http://www.chiantidatabase.org/.

CDMS -- Cologne Database for Molecular Spectroscopy, see website\\ 
http://www.ph1.uni-koeln.de/vorhersagen/, provides recommendations for
spectroscopic transition frequencies and intensities for atoms and
molecules of astronomical interest in the frequency range 0-10 THz, i.e.
0-340 cm$^{-1}$ (\cite{Mull05}).

BASECOL database (\cite{Dube06, Dube11b, Dube13}), see website
http://basecol.obspm.fr, is devoted to collisional rovibrational
excitation of molecules by atoms, ions, molecules or electrons.
It contains excitation rate coefficients for rovibrational excitation
of molecules by electrons, He and H$_2$ and it is mainly used for the 
study of interstellar, circumstellar and cometary atmospheres.

TIPTOPbase (http://cdsweb.u-strasbg.fr/topbase/home.html) contains:\\
 (i) TOPbase (\cite{Cunt92, Cunt93}), that lists atomic data computed in the 
Opacity and IRON Projects;
namely {\it LS}-coupling energy levels, {\it gf}-values and photoionization cross
sections for light elements (Z $\le$ 26) of astrophysical interest and \\
large datasets of {\it f}-values for ions of Fe with configurations 
3s$^x$3p$^y$3d$^z$. (ii) TIPbase (\cite{Naha03}) contains fine-structure 
atomic data computed for ions of astrophysical interest in the IRON Project:
radiative transition probabilities, electron impact excitation cross sections
and rates for fine-structure transitions. 

HITRAN -- (HIgh-resolution TRANsmission molecular absorption database) 
is at
\linebreak
http://www.cfa.harvard.edu/hitran/ (\cite{Roth09, Roth13, Hill13}). It lists
individual line parameters for molecules in the gas phase (microwave
through to the UV), photoabsorption cross-sections for many molecules,
and refractive indices of several atmospheric aerosols.  A high
temperature extension to HITRAN is the High-temperature molecular 
spectroscopic database - HITEMP (To access the HITEMP data: ftp to
cfa-ftp.harvard.edu; user = anonymous; password = e-mail address; 
cd/pub/HITEMP-2010).
It contains data for water, CO$_2$, CO, NO and OH (\cite{Roth10}).

GEISA -- (Gestion et Etude des Informations Spectroscopiques
Atmosph\'eriques) is a computer-accessible spectroscopic database, 
designed to facilitate accurate forward radiative transfer calculations
using a line-by-line and layer-by-layer approach. It can be found at
http://ether.ipsl.jussieu.fr/etherTypo/?id=950 (\cite{Jacq08, Jacq11}).
The current 2011 edition of GEISA is a system comprising three independent 
sub-databases devoted respectively to: (i) Line transition parameters,
(ii) Absorption cross-sections and (iii) Microphysical and optical 
properties of atmospheric aerosols.

NIST -- The National Institute of Standards and Technology hosts a
number of useful databases for Atomic and Molecular Physics.
A list can be found at\\
http://www.nist.gov/srd/atomic.cfm.  Among them are: An atomic spectra
database \\ (http://www.nist.gov/pml/data/asd.cfm) and three bibliographic 
databases providing references on atomic energy levels and spectra \\
(http://physics.nist.gov/cgi-bin/ASBib1/ELevBib.cgi), 
transition probabilities \\
 (http://physics.nist.gov/cgi-bin/ASBib1/TransProbBib.cgi)
and spectral line shapes and line broadening 
(http://physics.nist.gov/cgi-bin/ASBib1/LineBroadBib.cgi).

STARK-B database (http://stark-b.obspm.fr) contains theoretical widths 
and shifts of isolated lines of atoms and ions due to collisions with
charged perturbers, obtained using the impact approximation
(\cite{Saha10, Saha12, Saha14b}).

BEAMBD -- Belgrade electron/atom(molecule) database 
(\cite{Mari15} at http://servo.aob.rs/emol/). Contains 
electron interaction cross-sections for elastic scattering, electron 
excitation, ionization and total scattering.

The Virtual Atomic and Molecular Data Centre (VAMDC - http://www.vamdc.org/
- \cite{Dube11a, Rixo11})
provides the international research community with 
access to a broad range of atomic and molecular (A\&M) data compiled within 
a set of A\&M databases (including all the above mentioned except NIST)
accessible through the provision of a single portal.

\end{document}